\documentclass{PoS}
\usepackage{amsmath,amssymb}

\newcommand{\tmtextbf}[1]{{\bfseries{#1}}}

\newcommand{\tmtexttt}[1]{{\ttfamily{#1}}}

\newcommand{\onecol}[2]{
        \begin{minipage}[t]{#1}{#2\vfill} \end{minipage}
        }
\newcommand{\bea}{\begin{eqnarray}}
\newcommand{\eea}{\end{eqnarray}}
\newcommand{\eq}[1]{Eq.~(\ref{#1})}
\newcommand{\fig}[1]{Fig.~\ref{#1}}

\def\MSb{\overline{{\rm MS}}}
\def\Nf{N_{\rm f}}
\def\CF{C_{\rm F}}

\def\gc{\bar{g}_\mathrm{c}}

\newcommand{\ev}[1]{\left\langle #1 \right\rangle}
\newcommand{\Estat}{E_{\rm stat}}

\def\gqq{\bar{g}_\mathrm{qq}}
\def\aqqr{\alpha_{\rm qq}(1/r)}
\def\bg{\bar{g}}
\def\gs{\bg_{\rm S}}
\def\betas{\beta_{\rm S}}

\title{The shape of the static potential with dynamical fermions}

\ShortTitle{The shape of the static potential with dynamical fermions}

\author{
   \includegraphics[width=0.25\linewidth]{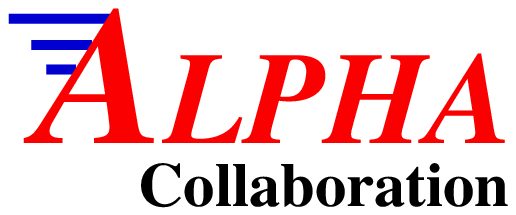}
   \hfill
   \onecol{3cm}{\vspace{-4.5em}\it
      BUW-SC 2011/08\\
      WUB/11-21
   }
   \vspace{1cm}
}
\author{\speaker{Bj\"orn Leder}$^{a,b}$ and
        Francesco Knechtli$^a$\\
        \llap{$^a$}Department of Physics, Bergische Universit\"at Wuppertal\\
                   Gaussstr. 20, D-42119 Wuppertal, Germany\\
        \llap{$^b$}Department of Mathematics, Bergische Universit\"at Wuppertal\\
                   Gaussstr. 20, D-42119 Wuppertal, Germany\\
        E-mail: \email{leder@physik.uni-wuppertal.de}
}

\abstract{We present the analysis of the static potential extracted from
  Wilson loops measured on CLS ensembles generated with Wilson gauge action
  and $N_f=2$ flavors of O$(a)$ improved Wilson quarks at three different
  lattice spacings and a range of quark masses. The shape of the static
  potential at distances well below the string breaking region is studied in
  terms of renormalized couplings derived from the static force and its
  derivative. We comment on the (im)possibility of extracting the Lambda
  parameter at our smallest lattice spacing $a=0.05$ fm. Finally we give an
  update on the scale determination through $r_0$.
}

\FullConference{XXIX International Symposium on Lattice Field Theory \\
                 July 10-16 2011\\
                 Squaw Valley, Lake Tahoe, California}

\begin{document}

\section{Introduction \label{s_intro}}

In this contribution we present an analysis of the static potential on
lattices generated with Wilson plaquette gauge action and $\Nf=2$ flavors of
non-perturbatively $\mathcal{O}(a)$ improved Wilson quarks. Periodic boundary
conditions are used for all fields apart from anti-periodic boundary
conditions for the fermions in time. The static potential is extracted from
measurements of the on-axis Wilson loops and can therefore only be determined
for distances smaller than the string breaking distance.

\vspace{-0.25cm}

\section{Method \label{s_method}}

\vspace{-0.25cm}

Our method to extract the static potential from HYP smeared Wilson loops has
been presented in \cite{Donnellan:2010mx} and is based on two steps. In the
first step all the gauge links are HYP smeared \cite{Hasenfratz:2001hp} using
the HYP2 parameters
\bea
\alpha_1=1.0\,,\quad \alpha_2=1.0\,,\quad \alpha_3=0.5 \,. \label{HYP2params}
\eea
As the time-like links are concerned, this is equivalent to the choice of a
static quark action. The binding energy of a static-light meson is
$\Estat \sim \frac{1}{a}e^{(1)}g_0^2+\ldots$
and the HYP2 choice of smearing parameters minimizes $e^{(1)}$
\cite{Della Morte:2005yc}.
The static energies $V_n(r)$ ($V_0\equiv V$ is the static potential) depends
on the static quark action. Their contribution to the
expectation values of Wilson loops $W(r,T)$ of size $r$ in one of the
spatial direction and size $T$ in the time-direction is given by
\bea
\ev{W(r,T)} & \sim &
\sum_n c_nc_n^* {\rm e}^{-V_n(r)(T-2a)} \quad \mbox{(with $N_t\to\infty$
time-slices)} \,.
\eea
The coefficients $c_n$ depend on the choice of the space-like links. In the
second step of our procedure we construct a variational basis for them. The
space-like links are smeared using $n_l$ iterations of {\em spatial} HYP
smearing 
\bea
\ev{W(r,t)} & \longrightarrow & C_{lm}(r,T)
\eea
We use a basis of $l=1,2,3$ operators and
the generalized eigenvalue method to extract $V_n$
\cite{Luscher:1990ck,Blossier:2009kd}.
The error analysis is based on the method of \cite{Wolff:2003sm} and we add
a tail to the autocorrelation function to account for the slow modes
\cite{Schaefer:2010hu}.
\begin{figure}\centering
  \resizebox{7cm}{!}{\includegraphics[angle=0]{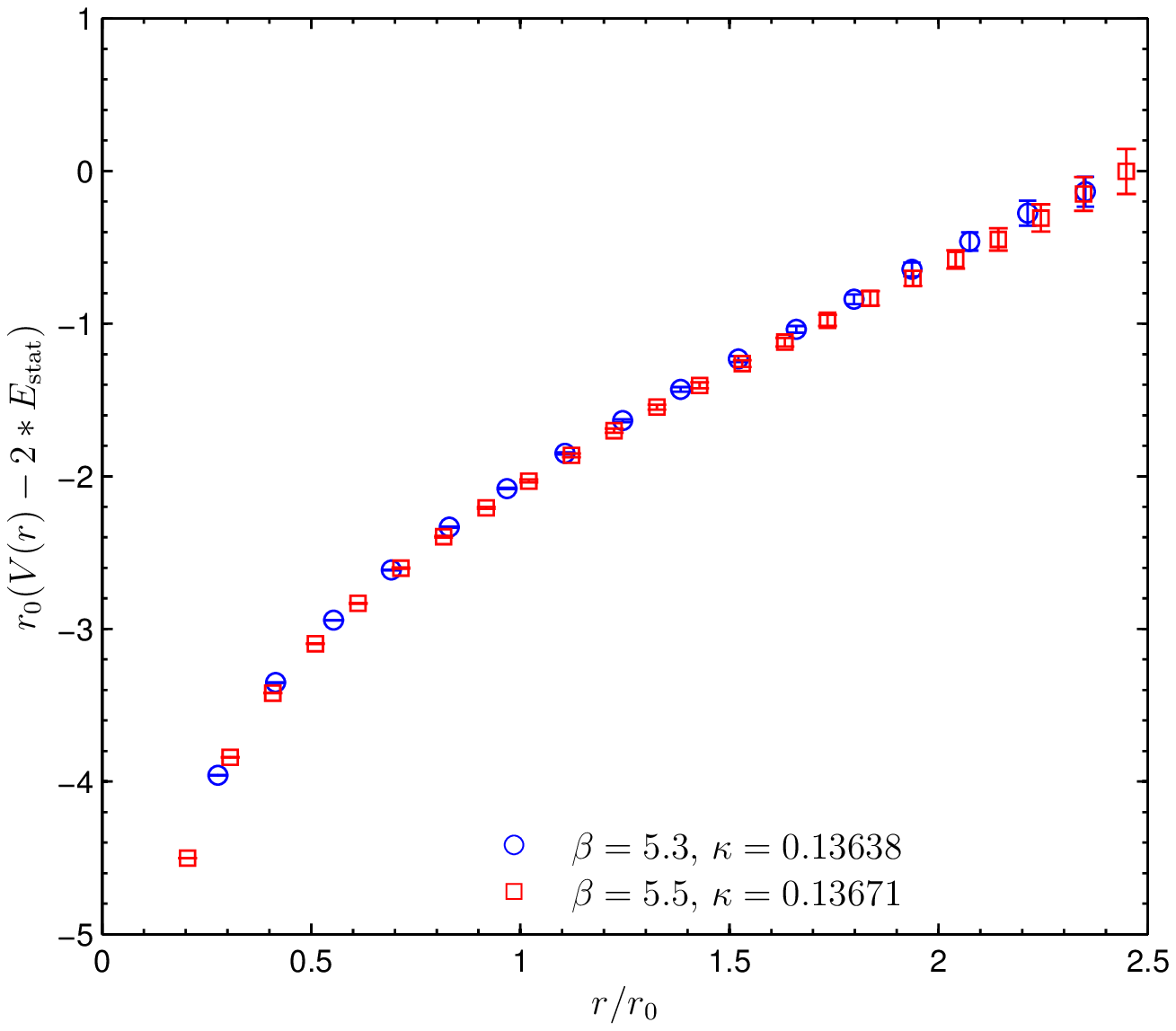}} \ \
  \resizebox{7cm}{!}{\includegraphics[angle=0]{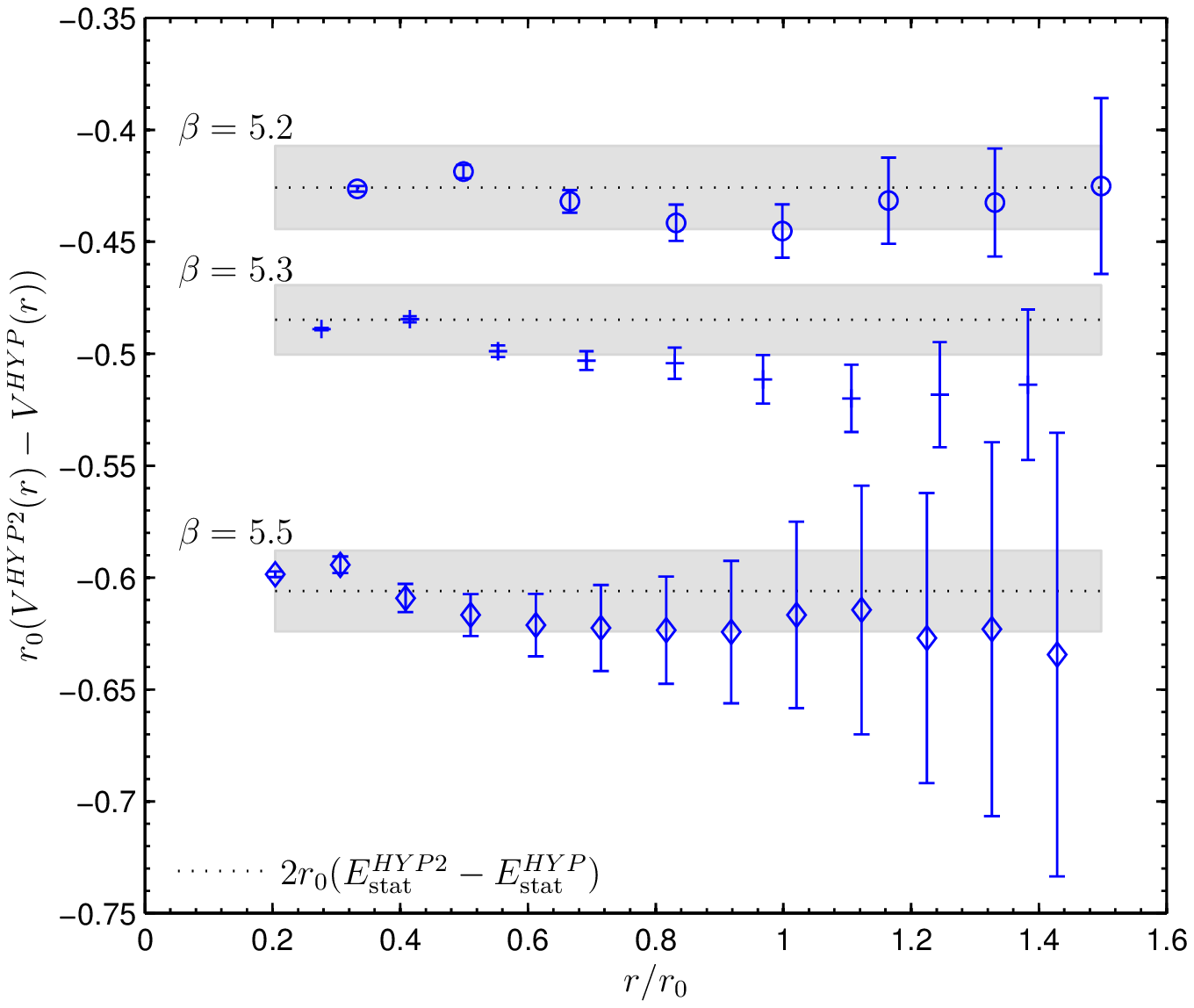}}
  \caption{Left plot: the static potentials on the F7a and O7 lattices. 
    Right plot: a comparison of the potentials using two static actions on the
    A5a, E5g and N5 lattices.}
  \label{f_pot}
\end{figure}

In this contribution we mainly present results for two lattice sets at
approximately constant pseudo-scalar mass $r_0M_{\rm PS}\approx0.62\ldots0.64$:
\bea
\mbox{F7a: 701 configurations}\,, \quad & 96\times48^3\,, & \quad r_0/a({\rm F7a})=7.079(63) \,,\\
\mbox{O7: 740 configurations}\,, \quad & 128\times64^3\,, & \quad r_0/a({\rm O7})=9.63(12) \,.
\eea
In the left plot of \fig{f_pot} we show the two potentials renormalized by
subtracting twice the energy of a static-light meson. The right plot of
\fig{f_pot} compares the potentials on the A5a, E5g and N5 lattices
\cite{preparation} computed with two different sets of HYP
parameters for the static action: the HYP2 set in \eq{HYP2params} and the set
called HYP \cite{Hasenfratz:2001hp}. The difference of the potentials is
\bea
r_0\,[V^{\rm HYP2}(r)-V^{\rm HYP}(r)] &=&
2\,r_0\,(\Estat^{\rm HYP2}-\Estat^{\rm HYP}) +
\frac{a^2\,r_0}{r^3}\,G(\Lambda r,m_qr) \,. \label{diffpot}
\eea
The dotted lines and gray bands represent the values with errors of 
$2\,r_0\,(\Estat^{\rm HYP2}-\Estat^{\rm HYP})$. We observe small cut-off
effects encoded in the function $G$ in \eq{diffpot}.

\vspace{-0.25cm}

\section{Scale $r_0/a$ \label{s_scale}}

\vspace{-0.25cm}

\begin{figure}\centering
  \resizebox{7cm}{!}{\includegraphics[angle=0]{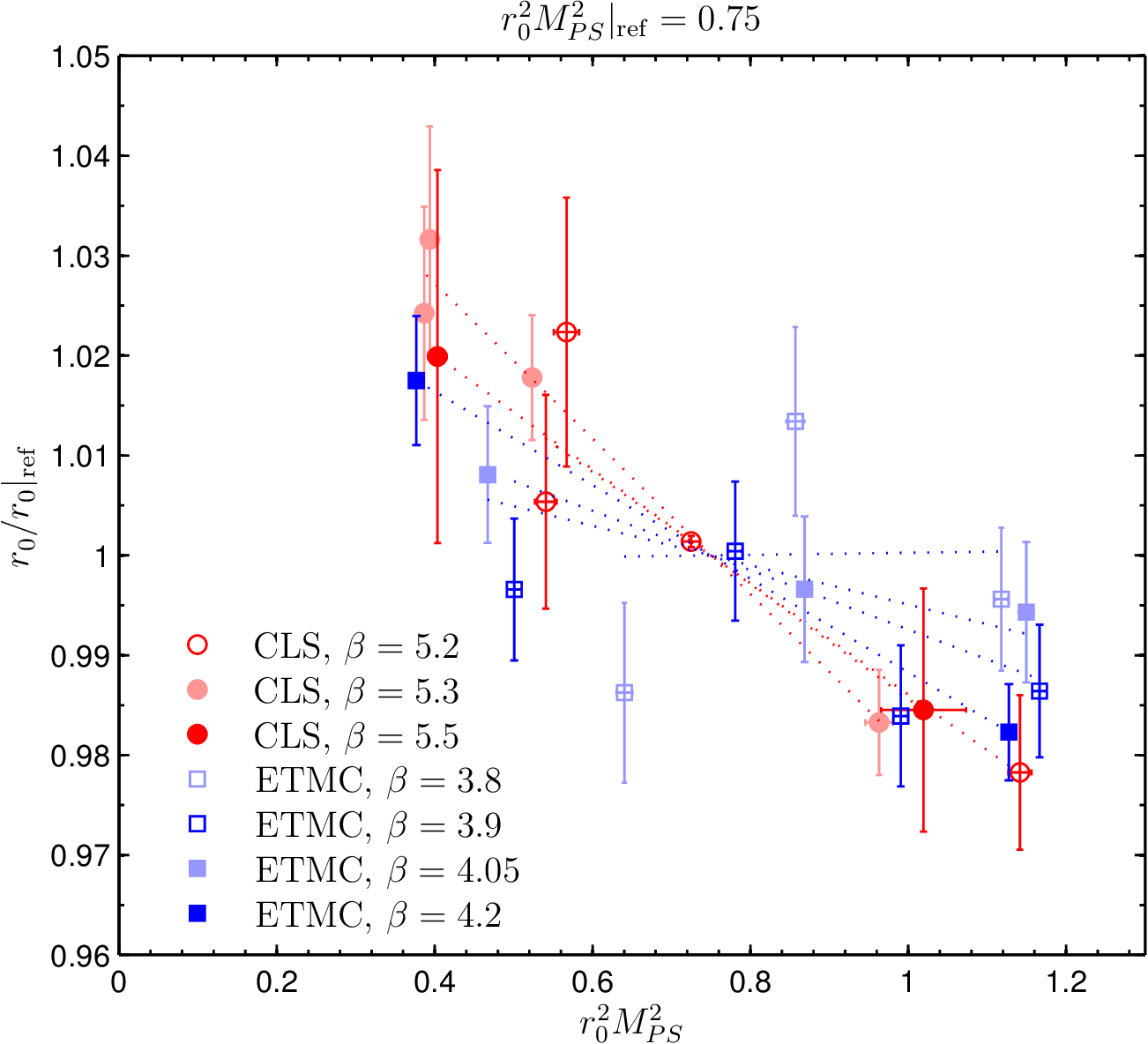}} \ \
  \resizebox{7cm}{!}{\includegraphics[angle=0]{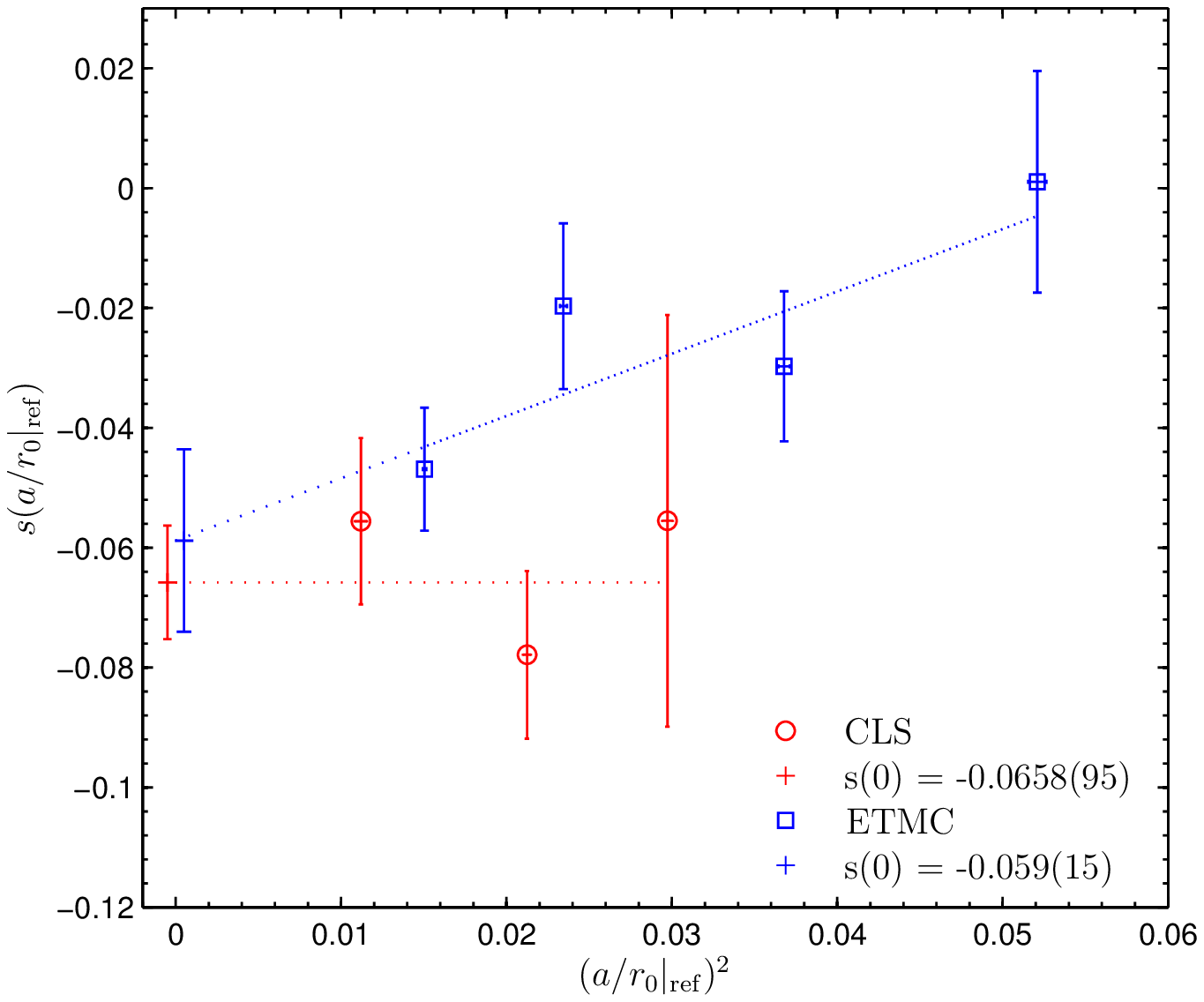}}
  \caption{Left plot: mass and cut-off dependence of $r_0$. Right plot:
    continuum extrapolation of the slope $s(a/r_0|_{\rm ref})$. Comparison
    with twisted mass data from \cite{Baron:2009wt}.}
  \label{f_sr0}
\end{figure}

Here we present a preliminary update on the determination of the scale $r_0$ 
\cite{Sommer:1993ce} at the three available $\beta$ values and based on the
currently available statistics. Motivated by the
right plot of fig. \ref{f_sr0} we perform a combined linear extrapolation
in the square of the pseudo-scalar mass to the chiral limit and obtain the
following results for $r_0/a(\beta)$:
\bea
r_0/a(5.2)=6.159(78)\,,\quad &
r_0/a(5.3)=7.242(70)\,,\quad &
r_0/a(5.5)=9.84(13)\,.
\eea
With these values of $r_0/a$ the preliminary update of the ALPHA
result for the $\Lambda$ parameter extracted from the running of the
Schr\"odinger Functional coupling \cite{DellaMorte:2004bc} is
\bea
r_0\,\Lambda_{\MSb}^{\Nf=2} & = & 0.78(3)(5) \,. \label{r0LambdaMSb}
\eea
The final numbers will be given in \cite{preparation}.
Alternative methods for determining the lattice spacing in the CLS consortium are
presented at this conference, through the kaon decay constant $F_{\rm K}$
\cite{marinkovic} or the $\Omega$ mass \cite{Capitani:2011fg}.

\vspace{-0.25cm}

\section{Mass dependence and cutoff effects in $r_0$ \label{s_cutoff}}

\vspace{-0.25cm}

In order to discuss the mass dependence and cut-off effects in $r_0$, we
introduce the variable $x = (r_0 M_{\rm PS})^2$, where $M_{\rm PS}$ is the
pseudo-scalar mass, and
define a reference value $r_0|_{\rm ref}$, which corresponds to the value of
$r_0$ at the pseudo-scalar mass
\bea
x_{\rm ref} \equiv (r_0 M_{\rm PS})^2|_{\rm ref} & = & 0.75 \,. \label{xref}
\eea
In the left plot of \fig{f_sr0} we plot $r_0/r_0|_{\rm ref}$ versus $x$
and compare our data with the twisted mass data of \cite{Baron:2009wt}. 
The value of $r_0|_{\rm ref}$ is obtained by linear interpolation and the
error analysis of $r_0/r_0|_{\rm ref}$ takes into account the correlations
between the data.
We only consider data with $x\le1.2$.

Using the data plotted in the left plot of \fig{f_sr0}, we determine for each
$\beta$ value the slope $s(a/r_0|_{\rm ref})$ in the Taylor expansion
\bea
\frac{r_0}{r_0|_{\rm ref}}(x) & = & 1 + s(a/r_0|_{\rm ref}) \cdot (x-x_{\rm ref})
\,.
\eea
The results are shown in the right plot of \fig{f_sr0}. In principle they
depend on the chosen value of $x_{\rm ref}=0.75$.
We repeated the analysis using $x_{\rm ref}=0.6$ but the
slopes do not change significantly. The twisted mass results for
$s(a/r_0|_{\rm ref})$ increase in magnitude as $a\to 0$ and they line up with our
results at the smallest $a$. Our data do not show significant cut-off
effects and we emphasize that the errors are larger due to accounting for the
slow modes in the autocorrelation function. In the continuum limit we obtain
the value $s(0)=-0.0658(95)$ by fitting to a constant.

\vspace{-0.25cm}

\section{Running couplings \label{s_couplings}}

\vspace{-0.25cm}

\begin{figure}\centering
  \resizebox{7cm}{!}{\includegraphics[angle=0]{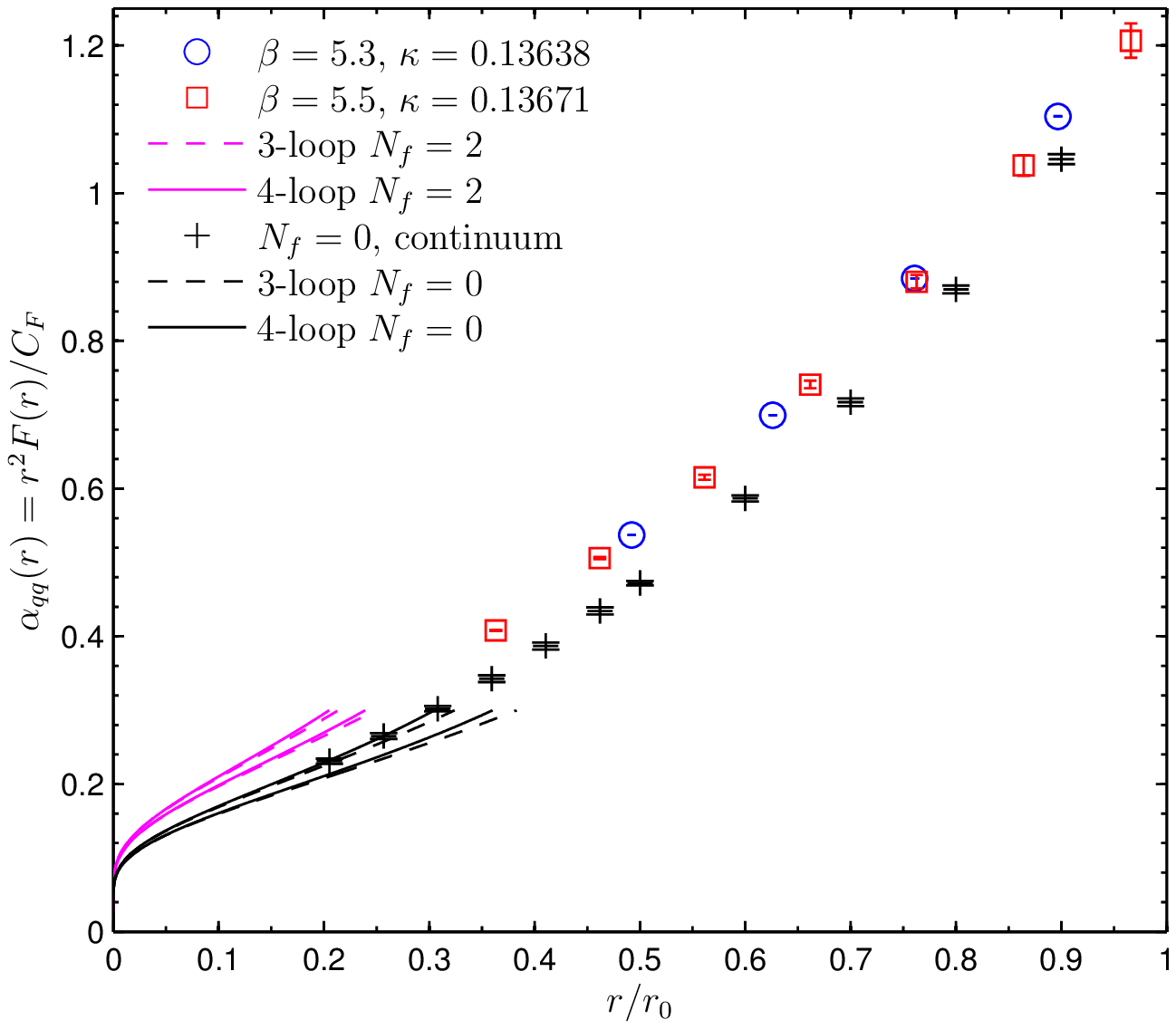}} \ \
  \resizebox{7cm}{!}{\includegraphics[angle=0]{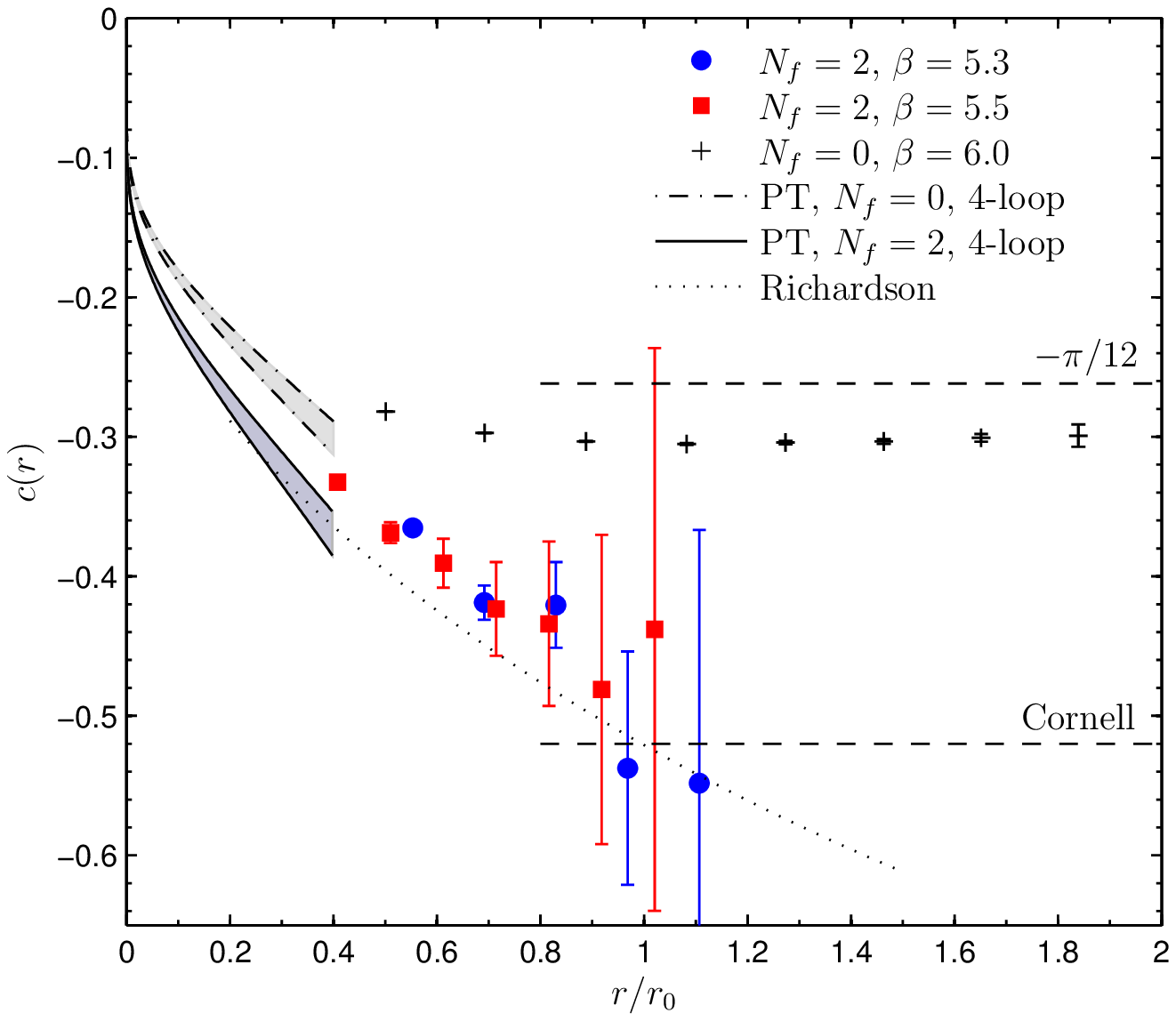}}
  \caption{Left plot: the coupling $\aqqr$ compared to perturbation theory and
    $\Nf=0$ results from \cite{Necco:2001xg}. Right plot: the slope $c(r)$
    compared to perturbation theory, potential models and $\Nf=0$ results from
    \cite{Luscher:2002qv}.}
  \label{f_couplings}
\end{figure}

\begin{figure}\centering
  \resizebox{9cm}{!}{\includegraphics[angle=0]{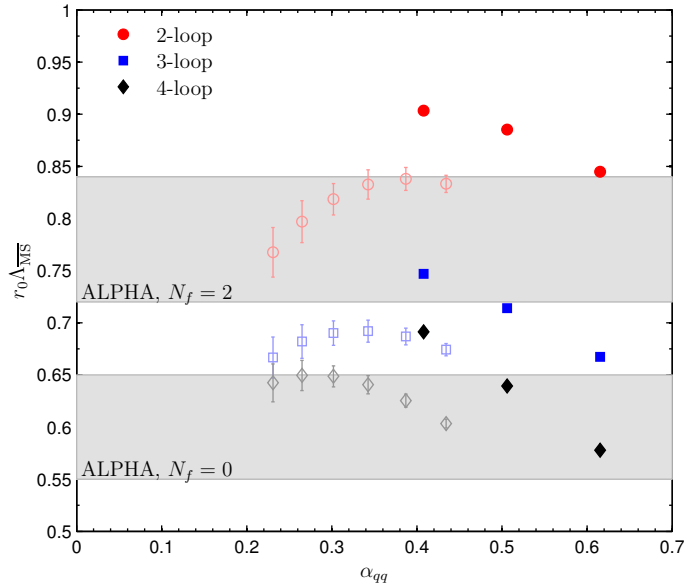}}
  \caption{Attempt to extract the $\Lambda$ parameter solving the RG equation
    for $\aqqr$ with the 2-, 3- and 4-loop $\beta$ function. $\Nf=0$ data are
    from \cite{Necco:2001xg}.}
  \label{f_Lambda}
\end{figure}

The static force $F(r)=V^\prime(r)$ defines the running coupling in the
qq-scheme $\gqq$ through
\bea
\gqq^2(\mu) &=& \frac{4\pi}{\CF}\,r^2\,F(r)\,,\;\mu=1/r \,.
\eea
We use an improved lattice definition of $F$ which removes the cut-off effect
effects at tree level \cite{Donnellan:2010mx}. The results for
$\aqqr=\gqq^2/(4\pi)$ on the F7a and O7 lattices are plotted in the left plot
of \fig{f_couplings}. The magenta dashed and continued lines are the 3- and
4-loop running respectively. They are obtained by solving the
renormalization group (RG) equation, which for a given scheme S reads (here
S=qq)
\bea
 \frac{\Lambda_{\rm S}}{\mu}  &=&
  \left(b_0\gs^2\right)^{-b_1/(2b_0^2)} {\rm e}^{-1/(2b_0\gs^2)}
           \exp \left\{-\int_0^{\gs} {\rm d}  x
          \left[\frac{1}{ \betas(x)}+\frac{1}{b_0x^3}-\frac{b_1}{b_0^2x}
          \right]
          \right\}\,. \label{RG}
\eea
For the beta function $\betas$ we use the 3- and 4-loop approximations
\cite{Brambilla:1999qa,Brambilla:1999xf,Smirnov:2009fh,Anzai:2009tm} and
$r_0\Lambda_{\rm S}$ is calculated from the value in \eq{r0LambdaMSb}
\cite{Donnellan:2010mx}. One sees that our smallest lattice spacing is
not small enough to make contact with perturbation theory. This becomes more
evident by the attempt to determine the $\Lambda$ parameter through \eq{RG}
shown in \fig{f_Lambda}. We use the $\aqqr$ values measured on the O7 lattices
as input for the right hand side of \eq{RG} together with the perturbative
beta function (2-loop, filled circles; 3-loop filled squares;
4-loop filled diamonds) and calculate $r_0\Lambda_{\MSb}$. For $\aqqr\to0$ the
values for the different loop approximations should converge to one value, but
this is not the case for our data in \fig{f_Lambda} (note that there is no
sign of the 2-loop data bending toward the 3- and 4-loop ones). For comparison
we plot the $\Nf=0$ data from \cite{Necco:2001xg} which demonstrate that
couplings down to at least $\aqqr\approx0.2$ (half the value of our smallest
coupling on the O7 lattice) would be needed.

From the slope
\bea
c(r) &=& \frac{1}{2}\,r^3\,F^\prime(r)
\eea
the running coupling $\gc$ is obtained
\bea
\gc^2(\mu) &=& -\frac{4\pi}{\CF}\,c(r)\,,\;\mu=1/r \,,
\eea
which defines the $c$-scheme. The results on the F7a and O7 lattices are
plotted in the right plot of \fig{f_couplings}. We also plot the $\Nf=0$
data from \cite{Luscher:2002qv} showing that there is a large effect
originating from dynamical fermions
(at distances much below the string breaking distance),
as already indicated by the perturbative curves.
For comparison we plot the values of $c(r)$ calculated using two
phenomenological potential models: the Cornell potential \cite{Eichten:1979ms}
\bea
V_{\rm Cl} & = & -\frac{\kappa}{r}+\sigma\,r \,,\quad \kappa=0.52
\eea
and the Richardson potential \cite{Richardson:1978bt}
\bea
V_{\rm R}(r) & = & \frac{1}{6\pi b_0}
\Lambda\left(\Lambda\,r-\frac{f(\Lambda\,r)}{\Lambda\,r}\right) \,.
\eea
If $r\,\Lambda\ll1$,
$V_{\rm R}(r)\,\sim\,-1/[6\pi b_0\,r\,\ln(1/\Lambda\,r)]$, where $b_0$ is the
1-loop coefficient of the beta function.
If $r\,\Lambda\gg1$,
$V_{\rm R}(r)\,\sim\,{\rm const}\times r$. Our data seem to follow the
Richardson curve (dotted line, \mbox{$\Nf=2$}, $r_0 \Lambda=0.73$) and approach the
Cornell value at distances
around $r_0$, where unfortunately the statistical errors (enhanced by the
factor $r^3$) become too large.

\vspace{-0.25cm}

\section{Conclusions}

\vspace{-0.25cm}

We determine the static potential from Wilson loops using the HYP2 static
action. Cut-off effects in the potential appear to be small and $r_0/a$ can be 
accurately determined.
There could be non-negligible cut-off effects in the mass dependence of
$r_0$ \cite{hep-lat/0309171},
however they seem to be small with improved Wilson fermions.

The running coupling from the static force is compared to
perturbation theory: at our lattice spacing we cannot reach small enough
distances in order to extract the $\Lambda$ parameter. This is in contrast to
what is concluded by a recent work with the same lattice spacing
\cite{Jansen:2011vv}. We observe large effects from dynamical fermions in the
slope $c(r)$. The shape of the static potential is relevant for holographic
QCD models \cite{Kol:2010fq,Giataganas:2011nz}.

\vspace{-0.25cm}

\section*{Acknowledgements}
\label{ack}
We thank Rainer Sommer for helpful discussions on various aspects of this work.
We further thank the Forschungszentrum J\"ulich and the Zuse Institute Berlin
for allocating computing resources to this project. Part of the Wilson loop
measurements were performed on the PC-cluster of DESY, Zeuthen.

\end{document}